# Can the Ferroelectric Soft Mode Trigger an Antiferromagnetic Phase Transition?

André Maia[1], Christelle Kadlec[1], Maxim Savinov[1], Rui Vilarinho[2], Joaquim Agostinho Moreira[2], Viktor Bovtun[1], Martin Kempa[1], Martin Míšek[1], Jiří Kaštil[1], Andriy Prokhorov[1], Jan Maňák[1], Alexei A. Belik[3] and Stanislav Kamba[1]*

[1]*Institute of Physics of the Czech Academy of Sciences, Na Slovance 1999/2, 182 21 Prague, Czech Republic*

[2]*IFIMUP, Physics and Astronomy Department, Faculty of Sciences, University of Porto, Rua do Campo Alegre s/n, 4169-007, Porto, Portugal*

[3]*International Center for Materials Nanoarchitectonics (WPI-MANA), National Institute for Materials Science (NIMS), Namiki 1-1, Tsukuba, Ibaraki 305-0044, Japan*

*e-mail: kamba@fzu.cz

Type-II multiferroics, where spin interactions induce a ferroelectric polarization, are interesting for new device functionalities due to large magnetoelectric coupling. We report on a new type of multiferroicity in the quadruple-perovskite $BiMn_3Cr_4O_{12}$, where an antiferromagnetic phase is induced by the structural change at the ferroelectric phase transition. The displacive nature of the ferroelectric phase transition at 125 K, with a crossover to an order-disorder mechanism, is evidenced by a polar soft phonon in the THz range and a central mode. Dielectric and pyroelectric studies show that the ferroelectric critical temperature corresponds to the previously reported Néel temperature of the $Cr^{3+}$ spins. An increase in ferroelectric polarization is observed below 48 K, coinciding with the Néel temperature of the $Mn^{3+}$ spins. This increase in polarization is attributed to an enhanced magnetoelectric coupling, as no change in the crystal symmetry below 48 K is detected from infrared and Raman spectra.

## 1. Introduction

Magnetoelectric multiferroics, where both magnetic and ferroelectric orders coexist and are coupled to one another, have received a lot of attention due to their potential numerous applications.[1,2] Quadruple perovskite multiferroics, with general chemical formula $(AA'_3)B_4O_{12}$, are now being extensively studied due to their intriguing and, in some cases, large magnetically-driven ferroelectric polarization.[3,4] Recently, magnetoelectric multiferroicity was found in the A-site ordered quadrupole perovskite $LaMn_3Cr_4O_{12}$ (LMCO), where two antiferromagnetic (AFM) phase transitions, associated with the G-type ordering of the A-site $Mn^{3+}$ and the B-site $Cr^{3+}$ spins, take place at 50 and 150 K, respectively.[5] Due to the interplay between the Mn and Cr magnetic sublattices, a ferroelectric polarization of $\sim 1.5 \times 10^{-3}$ µCcm$^{-2}$ emerges.[5] This novel type of multiferroicity has been explained by the anisotropic symmetric exchange mechanism.[6]

Since the $6s^2$ lone-pair electrons in ions like $Bi^{3+}$ can lead to a large ferroelectric polarization, as in the type-I multiferroic $BiFeO_3$ (60–100 µCcm$^{-2}$)[7], a new compound BiMn3Cr4O12 (BMCO) was recently sintered.[8] Surprisingly, BMCO was reported to exhibit both type-I and type-II multiferroic features.[8] Following the literature and starting from the high-temperature cubic phase with space group $Im\bar{3}$, BMCO undergoes the first ferroelectric phase transition (FE1 phase) at $T_{FE1} = 135$ K, marked by a broad anomaly in the temperature dependence of the permittivity ($\epsilon'_{max} \approx 420$) and the emergence of a spontaneous polarization[8] of $\approx 1.39$ µCcm$^{-2}$ (at 70 K). The system enters the type-I multiferroic state at $T_{N1} \approx 125$ K, due to the stabilization of the G-type AFM order of the B-site $Cr^{3+}$ spin sublattice, as revealed by neutron powder diffraction measurements.[8] On further cooling towards $T_{N2} \approx 48$ K, the second G-type AFM transition occurs for the A'-site $Mn^{3+}$ sublattice.[8] The exchange interaction between these two magnetic sublattices induces a strong magnetoelectric coupling, which enhances the ferroelectric polarization by $\approx 0.19$ µCcm$^{-2}$ below $T_{FE2} = T_{N2}$.[8] This led the authors of reference [8] to claim that BMCO sequentially exhibits type-I and type-II multiferroic features, something unique among all multiferroics.

Although both LMCO and BMCO exhibit the same G-type AFM structure in the $Mn^{3+}$ and $Cr^{3+}$ spin sublattices, the spin-driven contribution to the ferroelectric polarization at low temperatures in BMCO ($\sim 0.19$ µCcm$^{-2}$) is more than two orders of magnitude larger than that in LMCO ($\sim 1.5 \times 10^{-3}$ µCcm$^{-2}$).[5,8] With the appearance of G-type AFM ordering on both $Mn^{3+}$ and $Cr^{3+}$ magnetic sublattices, the polar $Cm$ structure (suggested by first-principle calculations) triggers the non-relativistic exchange striction mechanism, leading to a large spin-driven ferroelectric polarization.[9] While in BMCO the underlying mechanism for multiferroicity is exchange striction, in cubic $Im\bar{3}$ LMCO it is the weaker relativistic anisotropic symmetric exchange.[6,9]

First-principles calculations suggest that the ferroelectric phase transition at $T_{FE1}$ is driven by a soft polar optical phonon, which is predicted to exhibit a remarkable eigen-displacement of $Bi^{3+}$, inducing polarization and contributing as much as 93% to the lattice dielectric constant, due to the stereochemical activity of Bi $6s^2$ lone-pair electrons[9,10]. However, experimental evidence of the displacive nature of the ferroelectric phase transition is yet to be provided. Despite this, neutron powder, synchrotron X-ray and electron diffraction experiments did not reveal any deviations from the cubic symmetry, most likely due to the small lattice distortions below $T_{FE1}$.[8]

The optical soft mode is generally a dynamical origin of displacive ferroelectric phase transitions. In this case, the increase in permittivity near $T_C$ is caused by a softening, i.e., decrease in frequency, of one of the transversal polar phonons.[11] This can be understood from the Lyddane-Sachs-Teller relation:[12]

$$\frac{\epsilon_0}{\epsilon_\infty} = \prod_{j=1}^{n} \frac{\omega_{LOj}^2}{\omega_{TOj}^2} \quad (1)$$

describing the relation of the static permittivity, $\epsilon_0$, and the high-frequency (electronic) permittivity, $\epsilon_\infty$, with the $j$-th longitudinal, $\omega_{LOj}$, and transverse, $\omega_{TOj}$, phonon frequencies. Assuming $\epsilon_\infty$ and all the $\omega_{LOj}$ to be temperature independent and that the static permittivity, $\epsilon_0$, follows the Curie-Weiss law, one of the transverse phonon frequencies (usually the lowest frequency phonon $\omega_{TO1} = \omega_{SM}$) should follow the Cochran law:[11]

$$\omega_{SM}^2 = A(T - T_C) \quad (2)$$

where $A$ is the Cochran constant and $T_C$ is the critical Curie temperature. As the predicted polar soft mode is infrared (IR) active, IR and THz spectroscopies are the best tools for verifying the predicted phonon softening when approaching $T_{FE1}$.

In this paper, we demonstrate that the ferroelectric phase transition is driven by the softening of a polar phonon. Moreover, we show that $T_{FE1} = T_{N1}$, meaning that the ferroelectric transition triggers the magnetic ordering, which is novel among type-II multiferroics. Microwave (MW) and EPR spectra hint that a strong dielectric relaxation, i.e., a soft central mode, should exist in the microwave region, approximately 30 K above $T_{FE1}$. The ferroelectric phase transition is shown to be predominantly of the displacive type, with a weak order-disorder character near $T_{FE1}$. Raman spectroscopy reveals a spin-phonon coupling below 130 K. The observed phonon activities are compared with the predictions from factor-group analysis and first-principles calculations.[9,10]

## 2. Experimental Section

$BiMn_3Cr_4O_{12}$ ceramics were prepared from stoichiometric mixtures of $Bi_2O_3$, $Mn_2O_3$, and $Cr_2O_3$ at 6 - 7.5 GPa and about 1370 K for 1 hour in 6 mm diameter Au capsules. The high-pressure and high-temperature sintering was necessary for the ordering of the Mn and Cr in the A and B perovskite sites, respectively. Laboratory powder X-ray diffraction (XRD) data were taken at room temperature using a MiniFlex600 diffractometer with $CuK_\alpha$-radiation (2θ range of 10-80°, a step width of 0.02°, and a counting speed of 1 °min$^{-1}$). XRD data were analyzed by the Rietveld method with RIETAN-2000 program.[31] Weight fractions of impurities were estimated by RIETAN-2000 from refined scale factors. The best ceramics contained some impurities, specifically 1% of $Bi_2O_2CO_3$ and 1% $Cr_2O_3$, while others contained up to 1% of $Bi_2O_2CO_3$, 3% of $BiCrO_3$, and 8% $Cr_2O_3$. The X-ray diffraction pattern at room temperature and scanning electron microscope images are shown in Figure S1 and S2. For our investigations, we used the ceramics with a lower amount of impurities, while the others were only used to see the effect of impurities on the dielectric properties (see Figure S3). It should be mentioned that some of the secondary phases are magnetic.[32,33] However, they exhibit different critical temperatures than $BiMn_3Cr_4O_{12}$, and a qualitatively different temperature dependence of magnetic susceptibility (see Figure S4).

The low-frequency complex dielectric permittivity was measured from 10 to 300 K in the 1 kHz – 1 MHz frequency range using a HP4284A impedance analyzer together with a closed-cycle cryostat. The temperature rate and applied AC electric field was about 0.5 Kmin$^{-1}$ and 1 Vcm$^{-1}$, respectively. The experimental specimen was fabricated as a 700 μm thin plane-parallel polished plate. Au-Pd electrodes with a diameter of 1 mm were evaporated using a Bal-Tex SCD 050 sputter coater onto the principal faces of the plates. The contacts for applying the electric field were provided by silver wires fixed to t6he electrodes with silver paste. The thermally stimulated depolarization current (TSDC) measurements were done using the Electrometer High Resistance Meter KEITHLEY 617. The temperature dependence of the current density was measured in heating runs, after cooling the sample down to 10 K under several poling electric fields. The specific heat was measured using a Quantum design PPMS14.

MW resonance transmission measurements were performed using a composite dielectric resonator (DR), consisting of the base cylindrical DR and sample.[21,34] The disc-shaped sample without electrodes (1.8 mm thick, 5 mm in diameter) was placed on top of the base DR. The base and composite DRs were measured in the cylindrical shielding cavity using the transmission setup with a weak coupling by an Agilent E8364B network analyzer. The $S_{21}$ transmission coefficients were recorded in the frequency range near and below the main $TE_{01\delta}$ mode of the base DR (≈5.8 GHz), i.e., from 1 to 7 GHz during heating from 10 to 400 K with a temperature rate of 0.5 Kmin$^{-1}$ in a Janis closed cycle He cryostat. The temperature dependences of the $S_{21}$ maxima, corresponding to the mean resonance frequencies, were used to study temperature evolution of the main excitations (of dielectric or magnetic origin) in the GHz range. The $TE_{01\delta}$ resonance frequency, quality factor and insertion loss of the base DR, with and without the sample, were determined.

The EPR spectra were measured at X-band (9.384 GHz) in the 4 − 300 K temperature range using the Bruker E580 ELEXSYS spectrometer equipped with the ER 4122 SHQE Super X High-Q cavity. The samples were placed on quartz rods of 4 mm in diameter. The following experimental parameters were used during the EPR measurements: 1.5 mW microwave power, 100 kHz modulation frequency, 0.2 mT modulation amplitude and 60 ms conversion time.

THz complex transmittance was measured using a custom-made time-domain spectrometer powered by a Ti:sapphire femtosecond laser with 35-fs-long pulses centered at 800 nm. The system is based on coherent generation and subsequent coherent detection of ultrashort THz transients. The detection scheme consists of an electro-optic sampling of the electric field of the transients within a 1-mm-thick, (110)-oriented ZnTe crystal as a sensor.[35] This allows measuring the time profile of the THz transients transmitted through the sample. Low-temperature IR reflectivity measurements were performed using a Bruker IFS-113v Fourier-transform IR spectrometer equipped with a liquid-He-cooled Si bolometer (1.6 K) serving as a detector. For both the THz complex transmittance and IR reflectivity measurements, the temperature control was done through an Oxford Instruments Optistat optical continuous He-flow cryostats with mylar and polyethylene windows, respectively. To fit the IR and THz spectra, due to a significant TO-LO splitting, a generalized-oscillator model with the factorized form of the complex permittivity was used:

$$\epsilon(\omega) = \epsilon'(\omega) + i\epsilon''(\omega) = \epsilon_\infty \prod_j \frac{\omega_{LOj}^2 - \omega^2 + i\omega\gamma_{LOj}}{\omega_{TOj}^2 - \omega^2 + i\omega\gamma_{TOj}} \qquad (3)$$

where $\omega_{TOj}$ and $\omega_{LOj}$ denote the transverse and longitudinal frequencies of the *j*-th polar phonon, respectively, and $\gamma_{TOj}$ and $\gamma_{LOj}$ denote their corresponding damping constants. The high-frequency permittivity, $\epsilon_\infty$, resulting from electronic absorption processes, was obtained from the room-temperature frequency-independent reflectivity tail above the phonon frequencies and was assumed temperature independent. The dielectric strength, $\Delta\epsilon_j$, of the $j$-th mode is:[36]

$$\Delta\epsilon_j = \frac{\epsilon_\infty}{\omega_{TOj}^2} \frac{\prod_k \omega_{LOk}^2 - \omega_{TOj}^2}{\prod_{k \neq j} \omega_{TOk}^2 - \omega_{TOj}^2} \tag{4}$$

The complex permittivity, $\epsilon(\omega)$, is related to the reflectivity, $R(\omega)$, by:

$$R(\omega) = \left|\frac{\sqrt{\epsilon(\omega)} - 1}{\sqrt{\epsilon(\omega)} + 1}\right|^2 \tag{5}$$

Unpolarized Raman spectra were recorded using a Renishaw inVia Qontor spectrometer with a 532 nm linearly polarized diode-pumped laser and an edge filter. Measurements were done at fixed temperatures from 10 to 300 K using a THMS600 Linkam stage cooled by a nitrogen flow down to 80 K and a custom-made closed-cycle helium cryostat down to 10 K. The laser power (1.65 mW) was chosen adequately to prevent heating the sample. The wavenumber at a given temperature, $\omega(T)$, of each Raman mode is obtained by the best fit of the Raman spectra with a sum of damped oscillators:[37]

$$I(\omega, T) = [1 + n(\omega, T)]^{-1} \sum_j \frac{A_{0j}\Omega_{0j}^2\Gamma_{0j}\omega}{\left(\Omega_{0j}^2 - \omega^2\right)^2 + \Gamma_{0j}^2\omega^2} \tag{6}$$

where $n(\omega, T)$ is the Bose-Einstein factor and $A_{0j}, \Omega_{0j}, \Gamma_{0j}$ are the strength, wavenumber, and damping coefficient of the j-th oscillator, respectively. In the temperature range where no anomalous behavior is observed, the temperature dependence of the wavenumber of the phonon frequencies can be well described by the normal anharmonic temperature effect due to volume contraction as temperature decreases:[38]

$$\omega(T) = \omega_0 + C\left[1 + \frac{2}{e^x - 1}\right] \tag{7}$$

with $x \equiv \hbar\omega_0/2k_B T$ and where $\omega_0$ and C are model constants, $\hbar$ is the reduced Planck constant and $k_B$ is the Boltzmann constant.

## 3. Results and Discussion

### 3.1. Simultaneous Ferroelectric and Magnetic Phase Transitions

Figure 1a shows the temperature dependencies of the real part of dielectric permittivity, $\epsilon'(T)$, measured at different fixed frequencies (1 – 1000 kHz range), and of the specific heat divided by temperature, $C_p(T)/T$. The $\epsilon'(T)$ curve exhibits anomalous behavior, peaking at 138 K, which has been interpreted in current literature as a manifestation of the ferroelectric phase transition.[8] The anomaly at 138 K is unusually broad, and the temperature for which $\epsilon'(T)$ is maximum lies 13 K above the high temperature anomaly observed in the $C_p(T)$ curve, at 125 K, in good agreement with reference [8]. $C_p(T)$ exhibits another anomaly at $T_{N2} = 48$ K. The anomalies of specific heat at 125 K and 48 K have been attributed as manifestation of the G-type AFM ordering of the $Cr^{3+}$ and $Mn^{3+}$ spins sublattices, respectively, corroborated by magnetic neutron diffraction.[8] The discrepancy between the temperatures where $\epsilon'(T)$ and $C_p(T)$ display anomalous behavior is incomprehensible, since ferroelectricity is predicted to be stabilized by the condensation of a polar phonon and, so, the lattice contribution to the specific heat should be large. This discrepancy, along with the rather broad anomaly in $\epsilon'(T)$, hint more complex mechanisms underlying the ferroelectric and antiferromagnetic phase transitions in this temperature range.

For these reasons, we have undertaken a careful study of the thermally stimulated depolarizing current (TSDC) density, $J$. As a representative example, Figure 1b shows the temperature dependence of the $J(T)$ curve, measured with a heating rate of 0.5 K min$^{-1}$, after cooling the sample under a poling electric field of 0.7 kV cm$^{-1}$. The $J(T)$ curve displays two anomalies, a double maximum between 110 and 140 K, and another one at 32 K, with the shape and width of the highest temperature anomaly being unusual for a typical paraelectric-to-ferroelectric phase transition. The calculated electric polarization emerges below 138 K, reaching the value of 0.82 µC cm$^{-2}$ at 60 K and 0.9 µC cm$^{-2}$ at 10 K. Performing a similar TSDC density measurement after cooling with a poling field of 5.7 kV cm$^{-1}$ yields a polarization of 3.7 µC cm$^{-2}$ at 10 K (see Figure S5 of Supporting Information). Furthermore, the polarization can be reversed by reversing the poling electric field (see Figure S6).

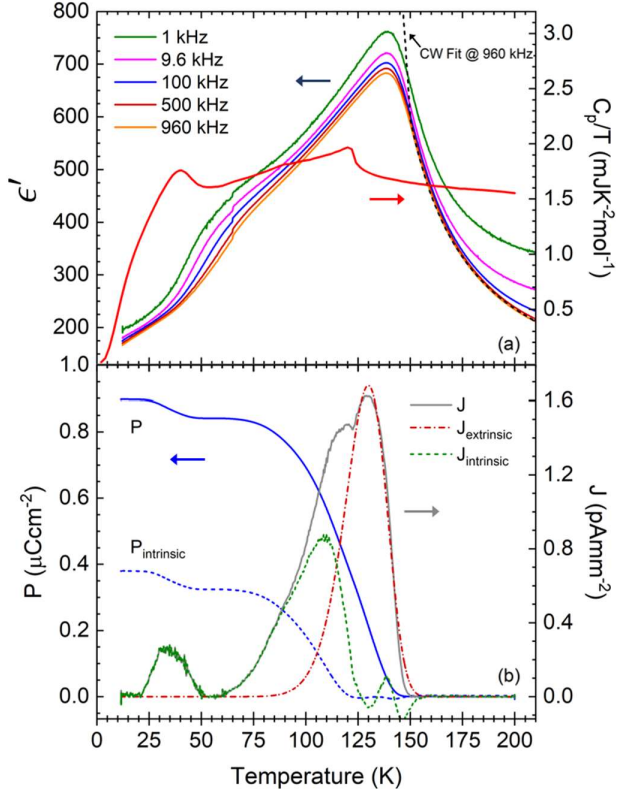

**Figure 1.** (a) Temperature dependence of the real part of permittivity measured at selected frequencies and of the specific heat divided by temperature. The Curie-Weiss fit of permittivity at 960 kHz is shown as a dashed line, giving $T_C \approx 125.8$ K (see main text). (b) Temperature dependence of the TSDC density (right axis), measured with a heating rate of 0.5 Kmin$^{-1}$ after cooling with a poling electric field of 0.7 kVcm$^{-1}$. The TSDC is decomposed by fitting the extrinsic component with Equation 8 in the 125−200 K temperature range (see main text). The intrinsic ferroelectric polarization, $P_{intrinsic}$, was calculated from the low-temperature (green) component, $J_{intrinsic}$. The total "polarization" $P$, including both intrinsic and extrinsic contributions, is plotted as a solid blue line.

The polarization values here reported are higher than the predicted values in reference [9], considering the polycrystalline character of our ceramics. Although the overall temperature dependence of permittivity and polarization qualitatively agree with those reported by Zhou et al.[8], quantitatively, we found higher values. The peak in permittivity at 138 K is almost 60% larger than in reference [8]. The enhanced permittivity seen in Figure 1a, mainly at low frequencies and above 138 K, is assigned to the Maxwell-Wagner effect, evidencing partial extrinsic contributions to both the electric permittivity and polarization, apparently enhancing both values. [20] In order to unravel the actual pyroelectric current, we measured the TSDC density at different heating rates (0.1 – 25 K min$^{-1}$, see Figure S7). The data recorded at the smallest temperature rate clearly shows two anomalies: a broader one peaking at 110 K and shaper one at 125 K. As the temperature rate increases, the higher temperature component upshifts by almost 25 K, while the lower one becomes sharper and the maxima shifts towards 128 K, and for high enough temperature rates (> 2 Kmin$^{-1}$) it becomes a shoulder around 138 K. The higher temperature TDSC peak is associated with extrinsic thermally stimulated depolarization processes (e.g., Maxwell-Wagner relaxation [13] or other defect-induced relaxations [14,15]), which have nothing to do with the ferroelectric depolarization process as we will show in the following. The TSDC density, $J$, recorded between 100 K and 150 K, can be decomposed into two contributions by modeling the extrinsic peak, hereafter designated by $J_{relaxation}(T)$, using the equation: [15]

$$J_{extrinsic}(T) \approx \frac{P_e}{\tau_0} exp\left(-\frac{E}{k_B T}\right) exp\left[-\frac{k_B T^2}{q\tau_0 E} exp\left(-\frac{E}{k_B T}\right)\right] \quad (8)$$

where $P_e$ is a constant, $\tau_0$ is the relaxation time at infinite temperature, $E$ is the activation energy of dipolar orientation, and $q = dT/dt$ is the heating rate. A detailed description of this model is given in reference [15] and a representative example of the best fit of Equation 8 to $J(T)$ is shown in Figure 1(b).

The intrinsic pyroelectric contribution to the TSDC density, $J_{intrinsic}$, is obtained by subtracting the extrinsic contribution from the total TSDC density, $J$, and its integration yields the actual ferroelectric polarization, $P_{intrinsic}(T)$ (Figure 1b). The polarization is slightly lower than in reference [8] due to the lower poling field used in this study. It should be noted that the polarization resulting from integrating $J_{intrinsic}$ becomes non-zero below 125 K, evidencing that the ferroelectric phase transition and the G-type AFM ordering of the $Cr^{3+}$ spin sublattice occur at the same temperature $T_{FE1} = T_{N1} = 125$ K. This result is consistent with the $C_p(T)/T$ curve, as it exhibits an anomaly at this temperature, as expected.

The stabilization of the ferroelectric phase below $T_{FE1}=125$ K is also evidenced by the hysteretic $P(E)$ loops, shown in Figure S8. Both the coercive electric field and remnant polarization increase as temperature decreases from 120 K down to 50 K. Below 50 K, the maximum applied electric field of 10 kVcm$^{-1}$ is not enough to induce a total domain reversal, and therefore, the measured hysteresis loops become narrower (see Figure S8). A similar effect has been reported in other ferroelectric systems with high electrical conductivity.[16,17]

The polarization increases below 50 K, i.e., approximately below $T_{N2}$ (see Figure 1b), most likely due to enhanced magnetoelectric coupling in the second AFM phase. Such anomaly in $P(T)$ does not necessarily require a change in crystal symmetry, as discussed in reference [8].

Furthermore, the dielectric dispersion seen near 50 K in Figure 1a is not typical of a ferroelectric phase transition. The relaxation frequency obtained from $\epsilon''(T)$ follows the Arrhenius behavior typical for the freezing of ferroelectric domain walls or other thermally activated process (see Figure S9 and accompanying discussion).

## 3.2. Displacive Nature of the Ferroelectric Phase Transition

Figure 2a and 2b show the real and imaginary parts of the complex dielectric spectra, $\epsilon'(\omega)$ and $\epsilon''(\omega)$, respectively, of BMCO measured at several fixed temperatures in the THz range. At room temperature, only one polar phonon is observed at 26 cm$^{-1}$. As temperature decreases from 300 K towards 125 K, the phonon wavenumber shifts towards lower frequencies, while the static electric permittivity increases, and on further cooling, the phonon hardens. This temperature behavior is typical for a polar soft phonon which drives the ferroelectric phase transition. Such behavior was predicted in reference [9].

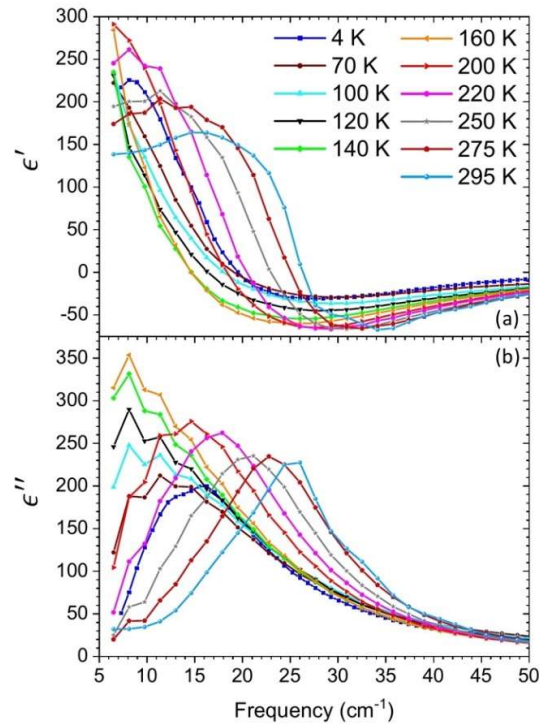

**Figure 2.** (a) $\epsilon'$ and (b) $\epsilon''$ spectra of BiMn$_3$Cr$_4$O$_{12}$ obtained by THz time-domain spectroscopy at several temperatures.

To better describe the temperature dependence of the polar soft phonon and its contribution to the total permittivity, we have also probed the higher energy polar lattice excitations through temperature dependent unpolarized FTIR spectroscopy. Representative reflectivity spectra, recorded at 4 and 250 K, are shown in Figure S10a. The spectra of the real and imaginary parts of permittivity, calculated from fitting the IR reflectivity spectra using Equation (3) and (5) (see Experimental Section), are also shown in Figure S10b and S10c, respectively. The parameters of polar phonons at 4 and 250 K are listed in Table SI, together with the predicted frequency values obtained from first-principles calculations in reference [10]. Three new polar modes activate in the ferroelectric phase, due to a change of symmetry from cubic $Im\bar{3}$ to the lower symmetry phase, which has the $Cm$ space group according to reference [9]. The factor-group analysis and the activity of phonons in IR and Raman spectra for both phases are also described in Supporting Information.

Figure 3a and 3b show the temperature dependence of the soft mode wavenumber and its dielectric strength, respectively. The soft mode frequency, $\omega_{SM}$, follows the Cochran law (Equation 2) above 180 K. Below this temperature, $\omega_{SM}(T)$ deviates from the theoretical expectation, most likely due to enhanced anharmonicity near the ferroelectric phase transition. The critical temperature obtained from the Cochran fit to the experimental data above 180 K is $T_{FE1} = (125 \pm 1)$ K, in excellent agreement with the intrinsic pyroelectric data and with the reported $T_{N1}$.

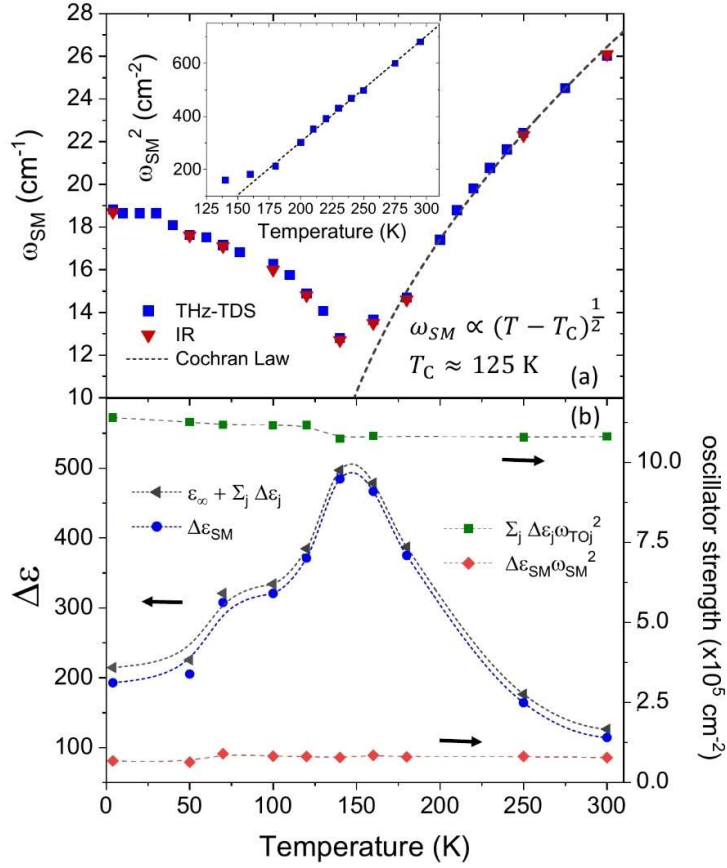

**Figure 3.** (a) Temperature dependence of the soft-mode frequency in $BiMn_3Cr_4O_{12}$. The squares and triangles are soft-mode frequencies from fits of THz time-domain and IR spectra, respectively. The dashed line in (a) was calculated from the best fit of Cochran law (Equation 2) to the soft mode frequency, in the $180 - 300$ K temperature range. (b) Contribution of the soft mode and the sum of the contributions of all phonons to the static dielectric permittivity (left axis) and temperature dependence of the soft mode oscillator strength and sum of all oscillator strengths (right axis). The dashed lines in (b) are a guide for the eye.

The dielectric strength of the soft phonon, $\Delta\epsilon_{SM}(T)$, considerably increases on cooling from 300 K towards $T_{FE1}$ (see Figure 3b). It is remarkable that $\Delta\epsilon_{SM}$ constitutes approximately 94% of the static electric permittivity in the THz range: $\epsilon_0 = \epsilon_\infty + \sum_j \Delta\epsilon_j$. It can also be seen in Figure 3b that the sum rule of oscillator strengths is verified, as $\sum_j \Delta\epsilon_j \omega_{TOj}^2$ is temperature independent within the accuracy of our measurements. The soft phonon oscillator strength, $\Delta\epsilon_{SM}\omega_{SM}^2$ is temperature independent, meaning that it is not significantly coupled with other phonons.[11] Based on these results, we have fitted the $\epsilon'(T)$ in the paraelectric phase using the Curie-Weiss law (revisit Figure 1a): $\epsilon'(T) = \epsilon_{HT} + C(T - T_C)^{-1}$, and we have obtained $T_C = (126 \pm 1)$ K, $C = 13327$ K and $\epsilon_{HT} = 83.6$. In this way, we again confirmed that the ferroelectric phase transition occurs ≈ 10 K below the peak in $\epsilon'(T)$. This seems unusual, but a similar behavior has been observed in relaxor-based ferroelectrics PZN-PT and PMN-PT, where some dielectric relaxation occurs above $T_C$.[18–20] In the next section, we describe the observation of a similar dielectric relaxation (central mode) in $BiMn_3Cr_4O_{12}$.

### 3.3. Displacive Ferroelectric Phase Transition with Order-Disorder Crossover

The deviation from the Cochran law ascertained below 180 K (55 K above $T_{FE1}$) in Figure 3a and the broad anomaly observed in $\epsilon'(T)$ around 135 K (see Figure 1a), are consequence of a dynamical central mode; i.e., a dielectric relaxation in the microwave range, which frequently activates near the Curie temperature in many displacive ferroelectrics due to the large lattice anharmonicity.[11] The static phonon permittivity (denoted by black triangles in Figure 3b) being 25% lower than the radio-frequency one (Figure 1a) supports this statement. To confirm this hypothesis, temperature dependent microwave resonance measurements were performed, and the corresponding results are shown in Figure 4a, revealing at least three temperature-dependent resonance modes above 150 K which are well seen in the frequency-temperature $S_{21}$ map.

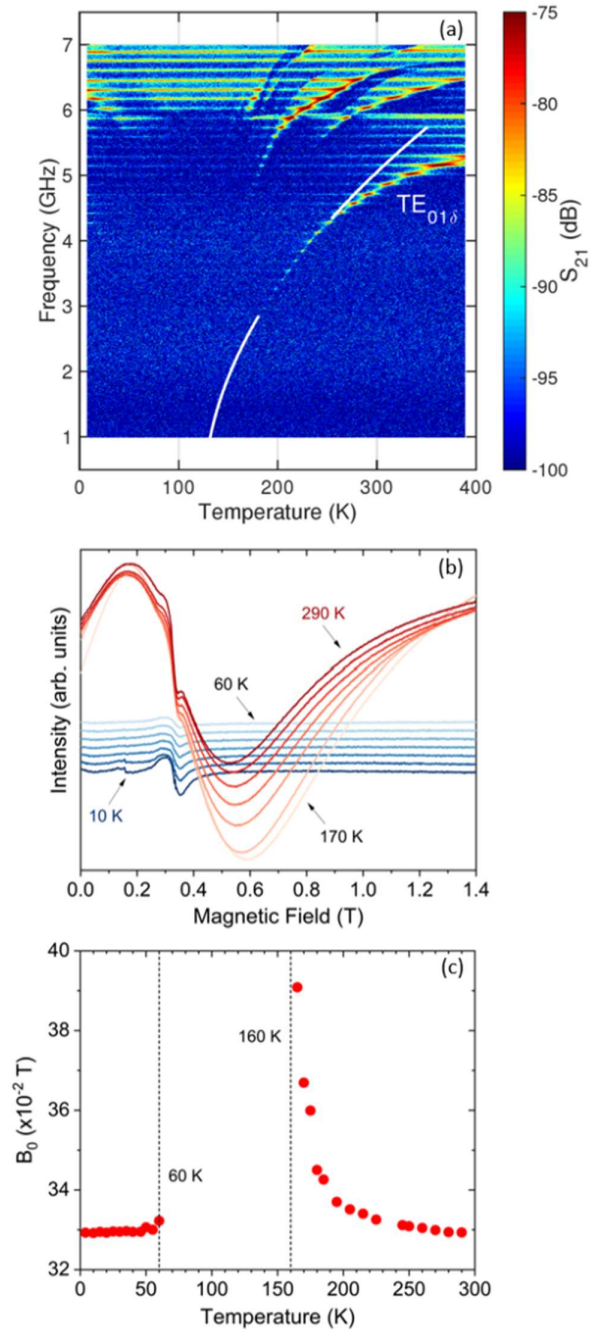

**Figure 4.** (a) Frequency-temperature map of $S_{21}$ transmission coefficient of the composite DR with $BiMn_3Cr_4O_{12}$. The colors correspond to $S_{21}$ values. The white solid line shows a fit of the $TE_{01\delta}$ mode frequency to a Cochran-like temperature dependence, as described in the main text. This line is discontinued between 180 and 260 K so that the experimental mode signal remains visible. (b) Derivative of the EPR spectra at 9.384 GHz between 4 and 290 K. The EPR spectra could not be measured between 60 and 160 K due to enhanced microwave losses caused by the central mode. (c) Temperature dependence of the magnetic field of the center line, $B_0$, which dramatically increases on cooling towards 160 K.

The lowest-frequency component is the $TE_{01\delta}$ mode of the composite dielectric resonator (DR). No spectral component could be seen below 150 K, due to the high microwave losses of BMCO, probably of both magnetic and polar origin. High microwave losses in the whole temperature range are evidenced by both high insertion loss (see $S_{21}$ scale in Figure 4a) and low quality factor $Q$ of the composite DR $TE_{01\delta}$ mode. The $Q$ factor of the base DR changes from 14 000 at 10 K, to 6000 at 390 K, while that of the composite DR ranges in the 200 – 300 interval, in the 190 – 390 K temperature range, and even less below 190 K, making resonances undetectable. This result indicates the existence of a dynamical central mode close to $T_{FE1}$. Despite the low signal-to-noise ratio, the $TE_{01\delta}$ mode resonance maxima are well defined in the $S_{21}$ spectra only above 160 K, allowing a reliable determination of the temperature dependence of the resonance frequency, $f_{TE}$. The $TE_{01\delta}$ mode slowing down on cooling is well described with the critical "Cochran-like" function: $f_{TE}(T) = f_o(T - T_c)^{1/2}$, with $f_0 = 382$ MHz and critical temperature $T_C \approx 125$ K. The critical behavior of $f_{TE}(T)$ resembles that of the soft phonon frequency, $\omega_{SM}(T)$, with the same $T_c$ (see Figure 3a). This is because the relation of the $TE_{01\delta}$ mode to the Curie-Weiss law (Figure 4a) with the microwave dielectric permittivity is similar to that of the soft mode, since $f_{TE}$ is mainly defined by the effective dielectric permittivity of the composite DR; i.e., by the electric permittivity of the base and sample.[21] Because the magnetic permeability of BMCO is $\mu' \neq 1$, the electromagnetic response, $\varepsilon'\mu'$, should be considered instead of electric permittivity.[22] As the electric permittivity of the base DR is nearly temperature independent, and the relative changes in temperature of the magnetic susceptibility of BMCO are much smaller than the ones in its electrical permittivity (see Figure S4 and reference [8]), the observed slowing down of the $TE_{01\delta}$ mode should be attributed to the Curie-Weiss law for microwave dielectric permittivity (Figure 4a), similarly to the soft mode. The deviation of $f_{TE}$ from the critical "Cochran-like" function at high temperatures is caused by the increasing influence of the temperature independent component of the composite DR effective permittivity, i.e., the dielectric permittivity of the base. Three other resonance modes (seen above $TE_{01\delta}$ in Figure 4a) also critically slow down on cooling. However, we cannot clearly define their type. Nevertheless, the critical temperature dependence of their frequencies above 160 K could be related to the Curie-Weiss law for the microwave dielectric permittivity of BMCO. At lower temperatures, these modes are overdamped by the high microwave magnetic and dielectric losses of the sample.

Figure 4b shows the derivative of EPR absorption line spectra of BMCO measured down to 4 K. The temperature dependence of the magnetic field, $B_0$, of the center line is shown in Figure 4c. The EPR spectra disappear below 160 K, 35 K above $T_{FE1}$, as the microwave dielectric loss strongly increases. The microwave loss decreases again below 60 K due to the slowing down of ferroelectric domain wall motion and the hardening of the optical soft mode. This makes it again possible to measure the EPR spectra at lower temperatures. The broadening of the EPR lines on cooling towards 160 K is a consequence of correlated atomic shifts in the local environment of the paramagnetic Mn and Cr centers occupying the A'- and B-sites.[16,23] At 48 K, there is a change in the resonant field $B_0$. This temperature roughly corresponds to $T_{N2}$.

The dramatic enhancement of the microwave loss below 160 K, evidenced by both microwave and EPR experiments, is explained by a dynamical central mode which is responsible for the deviation of the soft mode frequency from the Cochran law above $T_{FE1}$ (Figure 3a) and the fact that the phonon permittivity is 25% lower than the permittivity measured in the kHz frequency range (Figure 1a). This means that the ferroelectric phase transition in BMCO is displacive but with a crossover with order-disorder type close to $T_{FE1}$. Moreover, the dynamical central peak is also responsible for the broad anomaly observed in the low-frequency $\epsilon'(T)$ and explains why the maximum value occurs above to $T_{FE1}$.

### 3.4. Spin-Phonon Coupling

The spin-phonon coupling was investigated using Raman spectroscopy. The unpolarized Raman spectra recorded at several temperatures in the 100 – 800 cm$^{-1}$ range are displayed in Figure S11. The predicted Raman mode frequencies in the cubic phase, along with the experimental values observed at 10 K and 300 K, are shown in Table SII. Although expected from the symmetry point of view, no new bands could be observed below $T_{FE1}$ within the detection limit of the spectrometer. It can be explained by their overlapping with other modes or by weak intensities due to small lattice distortion in the ferroelectric phase. The later explanation is supported by results of X-ray and neutron diffraction, where no crystal symmetry change could be detected. [8]

However, a detailed inspection of the temperature dependence of some lattice modes reveals interesting features. Figure 5 shows the temperature dependence of selected Raman mode frequencies. The solid lines were obtained from the best fit of Equation 7 (see Experimental Section) to the experimental values in the temperature range $T > T_{FE1}$, extrapolated down to 0 K. The deviation from the expected anharmonic behavior makes it necessary to consider spin-phonon coupling. A frequency upshift (downshift) of a given phonon as a function of temperature due to the spin-phonon coupling should occur for dominant AFM (FM) interactions.[24,25] For the analyzed modes, the wavenumber follows the expected anharmonic temperature dependence from 300 K down to 125 K, but a downshift is seen below $T_{N1}$.

Interestingly, the same modes exhibit a frequency upshift in LMCO.[5] Even though no new bands were detected, the temperature dependence of the lowest frequency phonon (see Figure 5) exhibits a minimum at 75 K and below $T_{N2}$, it recovers the expected anharmonic temperature dependence of high temperature, extrapolated down to 10 K. This evidences structural changes at $T_{N2}$ induced by magnetism.

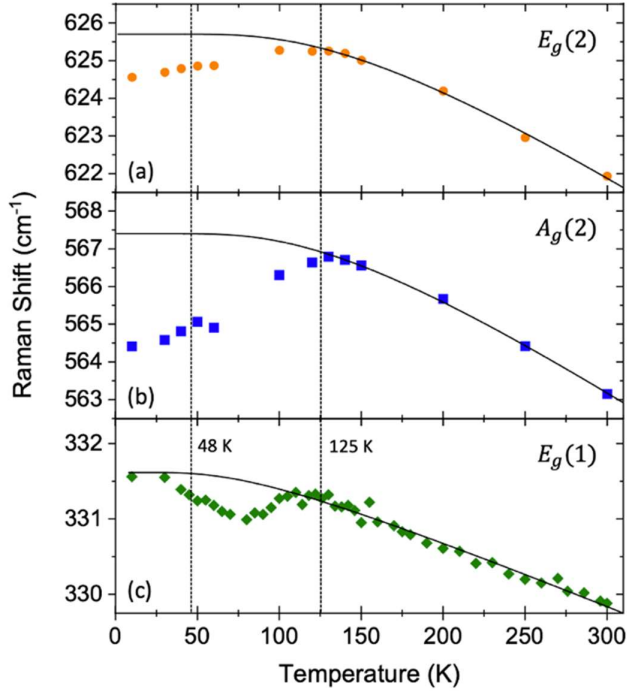

**Figure 5.** Temperature dependence of the frequencies of selected Raman modes. The best fit with Equation 7 in the paraelectric phase is represented by a solid line and is extrapolated down to 0 K.

The mechanisms underlying these lattice distortions are still unknown. These can be explained via the Dzyaloshinskii-Moriya interaction that, in non-centrosymmetric ferroelectric and AFM systems, such as BMCO, favors a spin canting of otherwise collinear (antiparallel) magnetic moments, thus acting as a source of weak ferromagnetic behavior in an antiferromagnet.[26] This is possible in BMCO because, unlike LMCO, it exhibits a polar space group below $T_{FE1}$. Another explanation could be the influence of the structural phase transition at $T_{FE1}$. Since the deviation of the phonon frequencies from the anharmonic behavior described by Equation 7 was observed also in LMCO [5], the spin-phonon coupling is a more plausible origin.

To obtain quantitative information regarding the order of the coupling term, we have analyzed the temperature dependence of the anomalous contribution to the phonon wavenumber, $\Delta\omega$, and the intensity of the magnetic neutron diffraction peaks, taken from reference [8]. The $\Delta\omega$ values were determined from the difference between the extrapolated anharmonic temperature dependence of the phonon wavenumber and the experimental one. The results are displayed in Figure 6.

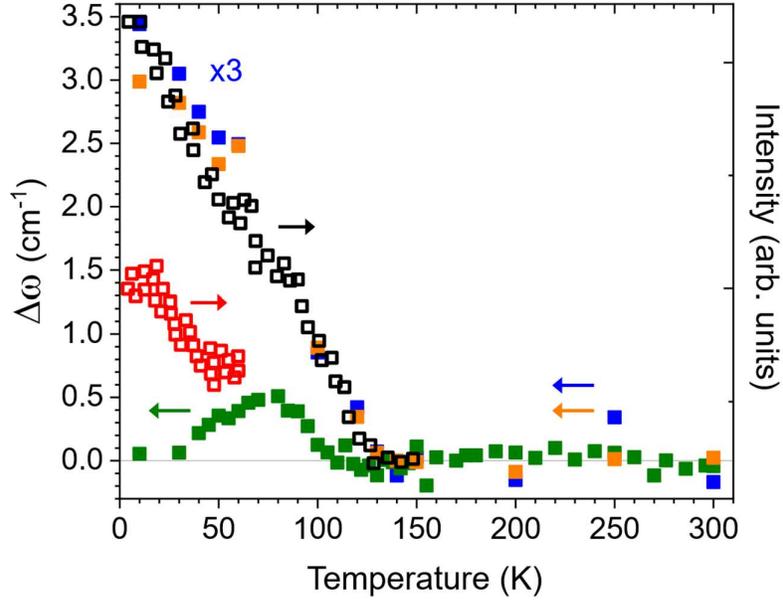

**Figure 6.** Temperature dependence of the anomalous contribution to the phonon wavenumber, Δω, (closed symbols, left axis) and the intensity of the magnetic neutron diffraction peaks (open symbols, right axis), taken from reference [8]. For comparison purposes, we have kept the same color code for the phonons as in Figure 5.

Considering the two highest wavenumber phonons shown in Figure 5, a similar temperature dependence to that observed for the magnetic peak intensity is found. As the intensity of the magnetic peak is proportional to the magnetic order parameters, a linear spin-phonon coupling is ascertained for these two phonons. The behavior of the lowest wavenumber phonon is interesting. Although it does not follow any magnetic peak diffraction temperature trend, the anomalous contribution Δω changes its temperature dependence at the same temperature as the magnetic peak associated with the magnetic phase transition taking place at $T_{N2}$. The results here reported clearly evidence an interplay between lattice and spin systems.

## 4. Conclusions

The experimental results here reported give clear evidence for the displacive nature of the ferroelectric phase transition, triggered by a polar soft phonon. Following the results reported by Dai and Zhang[9], the instability of the polar soft phonon, with $T_u$ symmetry in the cubic phase, mainly originates from the $Bi^{3+}$ cation. However, the $Bi^{3+}$ cation is also involved on the dynamical central peak, resulting in the ferroelectric phase transition having both displacive and order-disorder natures. This interpretation is supported by the results of Zhou et. al.,[8] as the temperature dependence of the thermal parameter of $Bi^{3+}$ shows a decrease below $T_{FE1}$. The existence of the dynamical central peak in the microwave spectral region explains the rather broad anomaly observed in the temperature dependence of the real part of the complex electric permittivity, around $T_{FE1}$.

The analysis of the temperature dependence of the TSDC density enables to unravel the extrinsic mechanisms for polarization and subtract their contribution to the overall polarization. Moreover, the ferroelectric critical temperature, $T_{FE1}$, extrapolated from the Cochran and Curie-Weiss laws, matches the Néel temperature, $T_{N1} = 125$ K. Based in these results, along with the clear anomaly observed in the specific heat as a function of temperature, we propose that both the ferroelectric phase transition and the G-type AFM ordering of the B-site $Cr^{3+}$ spins occur at the same temperature, i.e., $T_{FE1} = T_{N1} = 125$ K. Moreover, since the ferroelectric polarization is induced by the polar soft phonon, we consider polarization as the primary order parameter, which is coupled to the magnetic one. In this picture, the ferroelectric phase transition triggers the antiferromagnetic order in an originally paramagnetic/paraelectric state. Supporting this interpretation, we must consider the linear relationship between the anomalous contribution to the Raman-active phonon wavenumber and the intensity of the magnetic diffraction peaks, which clearly evidence a linear spin-lattice coupling in this compound. This interpretation makes $BiMn_3Cr_4O_{12}$ a special case of multiferroicity, although no microscopic theory exists yet. To the knowledge of the authors, there are only two theoretical papers[27,28] proposing that ferroelectricity can induce weak ferromagnetism in an originally antiferromagnetic phase, which is not applicable to this case. Nevertheless, one can expect that the structural change at $T_{FE1}$ can modify the magnetic exchange interactions, giving rise to an AFM order at the same temperature. A phenomenological theory of ferroelectric phase transitions triggered by an instability of the crystal with respect to another phase transition parameter has already been published by Holakovský.[29] A similar theory can be applied to an AFM phase transition triggered by a structural (ferroelectric) phase transition.

Below $T_{N2}$ = 48 K, the G-type antiferromagnetic ordering of Mn$^{3+}$ spin magnetic sublattice enhances the ferroelectric polarization. Both IR and Raman spectra do not disclose new bands pointing for any additional symmetry-breaking below that temperature. A similar enhancement of ferroelectric polarization has been reported in other multiferroics like BiFeO$_3$ and CdMn$_7$O$_{12}$.[4,30] However, precursor structural distortions occur above $T_{N2}$, as ascertained from the minimum value reached by the Raman-active phonon wavenumber, peaking at 330 cm$^{-1}$ (see Figure 5 and 6).

Summarizing, we reported an updated view of the phase sequence of BiMn$_3$Cr$_4$O$_{12}$. The possibility of a ferroelectric soft phonon instability triggering a G-type antiferromagnetic phase transition is proposed based on a comprehensive experimental study of this material. This is a remarkable result that will motivate a discussion on the coupling between the polar lattice instability and the magnetic order parameter.

**Acknowledgments**

This work has been supported by the Czech Science Foundation (Project No. 21-06802S) and the MŠMT Project SOLID 21- CZ.02.1.01/0.0/0.0/16_019/0000760. Experiments were partly performed in MGML (mgml.eu), which is supported within the program of Czech Research Infrastructures (project no. LM2018096).

**Competing interests**

The authors declare no competing interests.

# Supporting Information

## Can the Ferroelectric Soft Mode Trigger an Antiferromagnetic Phase Transition?

André Maia[1], Christelle Kadlec[1], Maxim Savinov[1], Rui Vilarinho[2], Joaquim Agostinho Moreira[2], Viktor Bovtun[1], Martin Kempa[1], Martin Míšek[1], Jiří Kaštil[1], Andriy Prokhorov[1], Jan Maňák[1], Alexei A. Belik[3] and Stanislav Kamba[1]*

*[1]Institute of Physics of the Czech Academy of Sciences, Na Slovance 1999/2, 182 21 Prague, Czech Republic*

*[2]IFIMUP, Physics and Astronomy Department, Faculty of Sciences, University of Porto, Rua do Campo Alegre s/n, 4169-007, Porto, Portugal*

*[3]International Center for Materials Nanoarchitectonics (WPI-MANA), National Institute for Materials Science (NIMS), Namiki 1-1, Tsukuba, Ibaraki 305-0044, Japan*

*\*e-mail: kamba@fzu.cz*

**Sample Characterization**

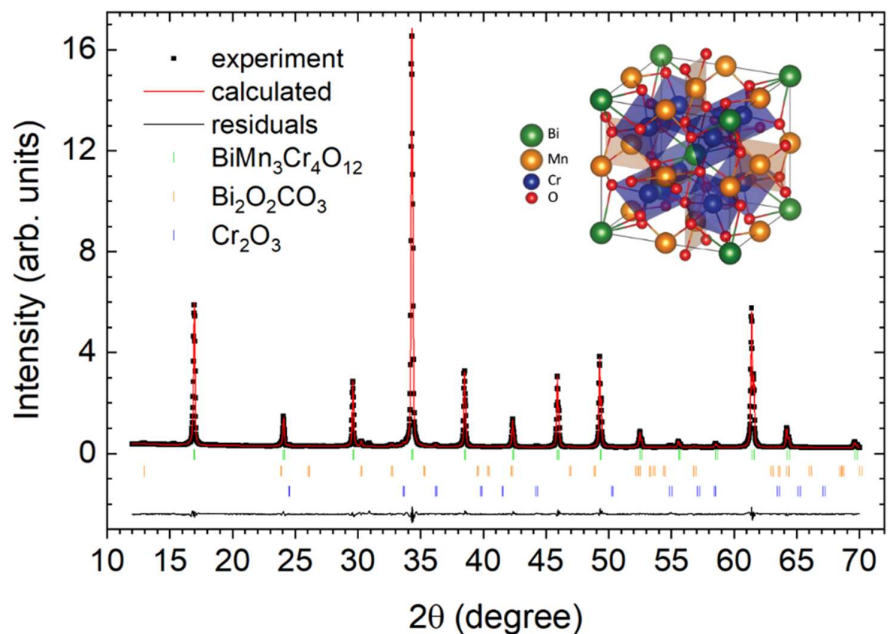

**Figure S1.** X-ray diffraction pattern and result of Rietveld refinements at room temperature. The cubic quadruple-perovskite structure of $BiMn_3Cr_4O_{12}$ is depicted in inset, made using VESTA software.[1] In this sample, 2% of $Bi_2O_2CO_3$ and 3% of $Cr_2O_3$ impurities were detected.

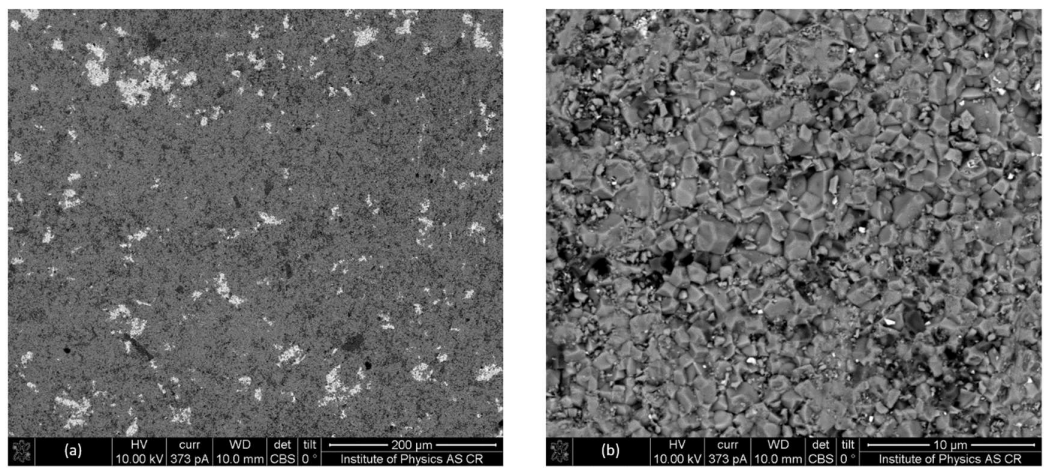

**Figure S2.** Representative SEM images of a $BiMn_3Cr_4O_{12}$ ceramic sample at room temperature. Three different phases can be observed. The most predominant one (grey regions) is $BiMn_3Cr_4O_{12}$, while the white and black regions correspond to $BiCrO_3$ and $Cr_2O_3$ impurities, respectively. The chemical composition of some of these regions was determined using EDS. The $Bi_2O_2CO_3$ secondary phase observed in XRD was not detected in EDS, as this ceramic was from another batch than the sample in Figure S1.

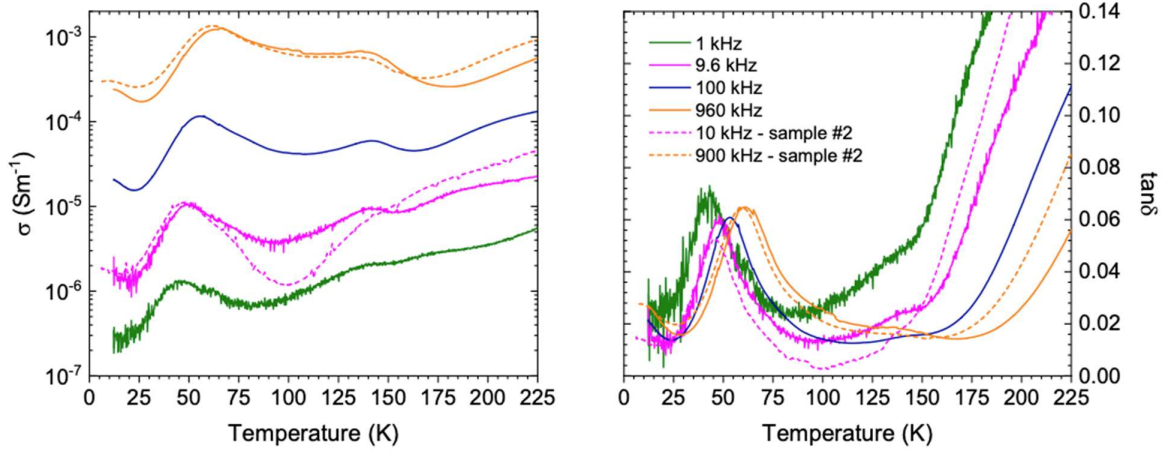

**Figure S3.** Temperature dependence of the conductivity and the dielectric loss measured at several frequencies. The dashed lines show results obtained at 9.6 kHz and 960 kHz on another sample (#2) with a higher impurity content (1% of $Bi_2O_2CO_3$, 3% of $BiCrO_3$, and 8% $Cr_2O_3$). The higher quality sample has 1% of $Bi_2O_2CO_3$, 1% $BiCrO_3$, and 3% $Cr_2O_3$.

**Factor group analysis of the lattice vibrations in the high-temperature cubic phase**

In the cubic paraelectric $Im\bar{3}$ phase of $BiMn_3Cr_4O_{12}$, the Bi, Mn, Cr and O ions occupy the $2a(0,0,0)$, $6b(0.5,0.5,0)$, $8c(0.25,0.25,0.25)$ and $24g(x,y,0)$ Wyckoff positions, respectively.[2, 3] The factor-group analysis of the $\Gamma$-point optical phonons gives the following irreducible representation: $11F_u \oplus 2E_u \oplus 2A_u \oplus 2E_g \oplus 2A_g \oplus 4F_g$.[4, 5] The $F_u$ modes are IR-active, the $E_g$, $F_g$ and $A_g$ modes are Raman-active, and the remaining $E_u$ and $A_u$ modes are silent. In the ferroelectric phase, assuming that BMCO crystallizes in the $Cm$ space group, the Bi, Mn and 4 oxygen ions are in the $2a(x,0,z)$ Wyckoff position, while the Cr and the remaining 8 oxygen ions occupy the $4b(x,y,z)$ one.[3] Therefore, the optical modes have symmetries $15A' \oplus 12A''$ and are both IR- and Raman-active.[4, 5] There is a group-subgroup relation between the $T_h$ and $C_s$ point groups of the paraelectric and ferroelectric phases, and therefore, a continuous ferroelectric phase transition is possible.[6] The proper ferroelectric polarization lies along the z-axis, the improper ferroelastic deformations $u_{xx}$, $u_{yy}$ and $u_{zz}$ are allowed and 6 types of ferroelectric domains are possible.[6]

**Magnetic Studies**

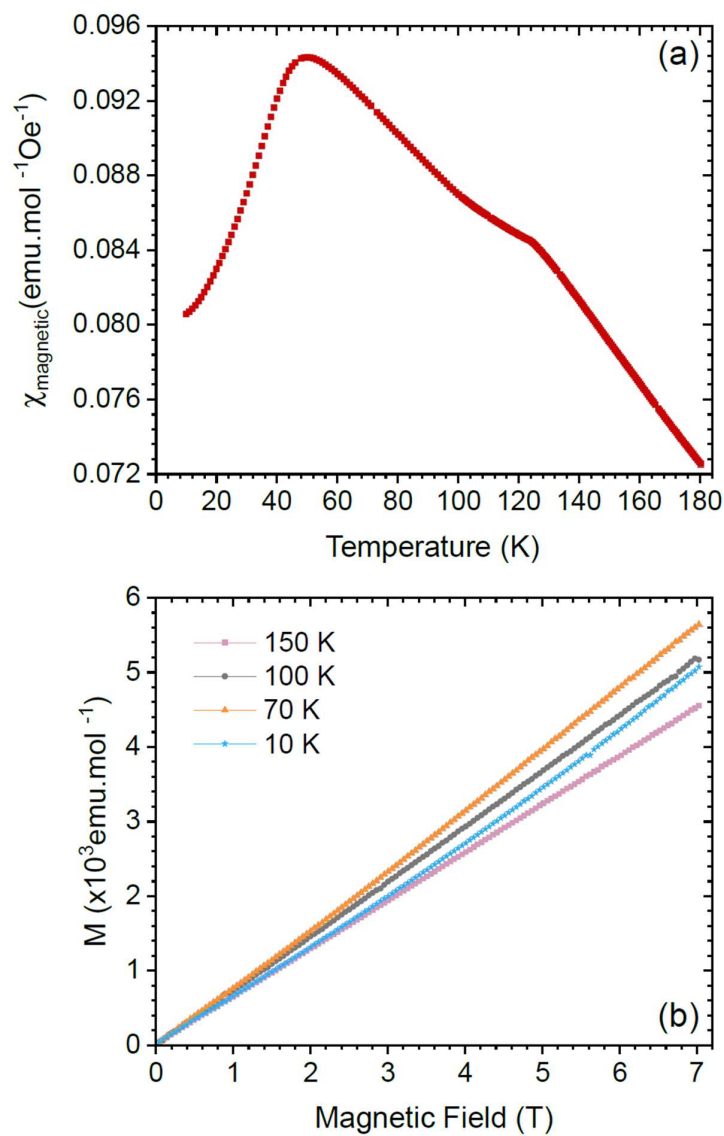

**Figure S4.** (a) Temperature dependence of the magnetic susceptibility of BiMn$_3$Cr$_4$O$_{12}$, measured under an applied magnetic field of 1 T. (b) Magnetic field dependence of the magnetization, measured at several temperatures. Magnetization is linearly dependent on magnetic field roughly up to 2 T, therefore $\chi_{magnetic} = M/H$ in that range.

# Polar and Dielectric Studies

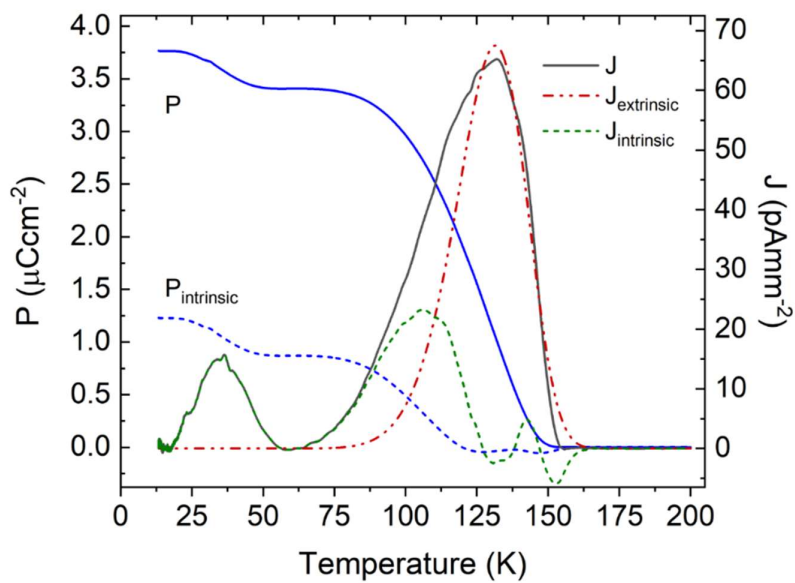

**Figure S5.** Temperature dependence of the thermally stimulated depolarization current (TSDC) density (right axis), measured with a heating rate of 0.5 Kmin$^{-1}$ after cooling with a poling electric field of 5.7 kVcm$^{-1}$. The TSDC is decomposed by fitting the extrinsic component using Equation 3 in the 125−200 K temperature range (see main text). The intrinsic ferroelectric polarization, $P_{intrinsic}$, was calculated by integrating in time the low-temperature (green) component, $J_{intrinsic}$. The total "polarization" $P$, including both intrinsic and extrinsic contributions, is plotted by solid blue line.

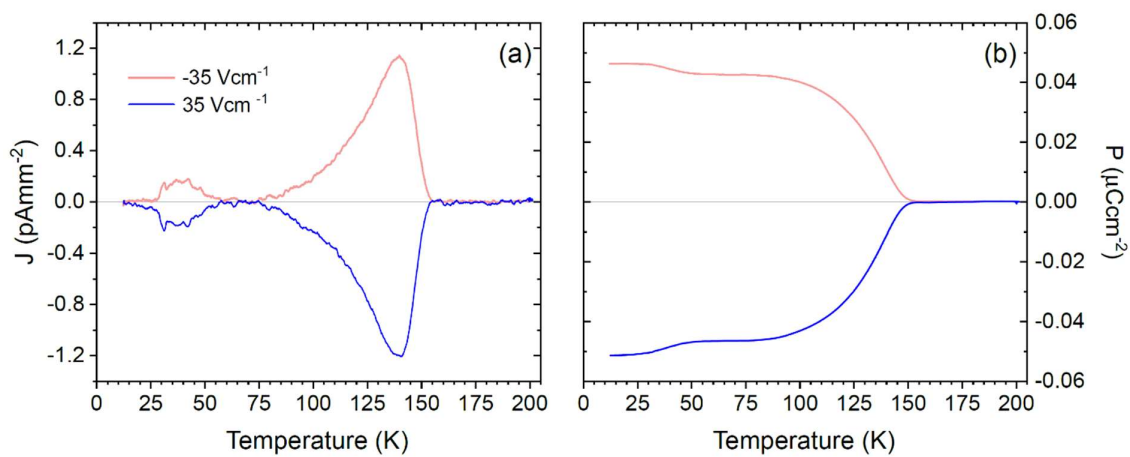

**Figure S6.** Temperature dependence of (a) pyro-current density and (b) polarization measured after poling with electric fields of ±35 V/cm, applied on cooling below 200 K. Note that even a small electric field can partially switch the polarization.

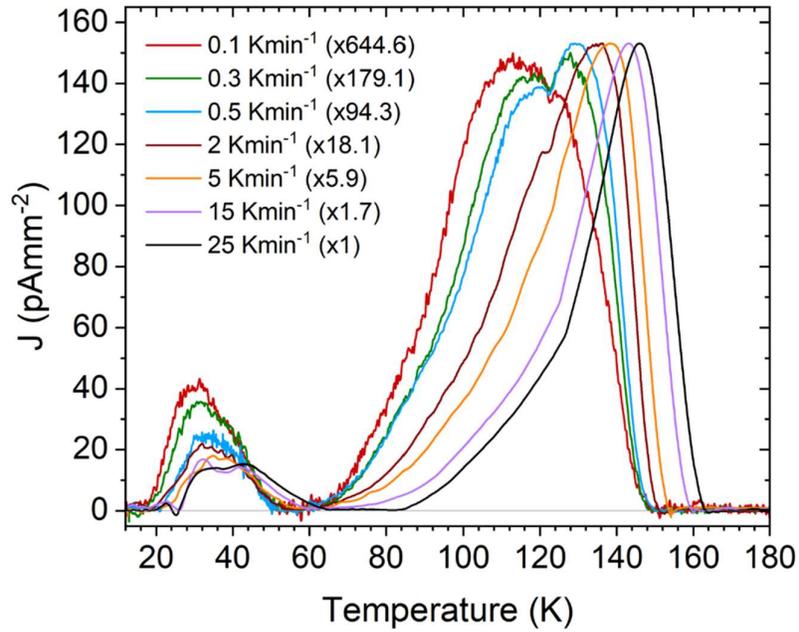

**Figure S7.** Temperature dependence of the TSDC density of $BiMn_3Cr_4O_{12}$ measured at several heating rates with a poling electric field of 0.7 kVcm$^{-1}$. For clarity, the curves were each multiplied by the factor indicated in the legend. The odd shape of the peak near 40 K in the current measured at 25 Kmin$^{-1}$ is caused by instabilities in the temperature ramp below 60 K.

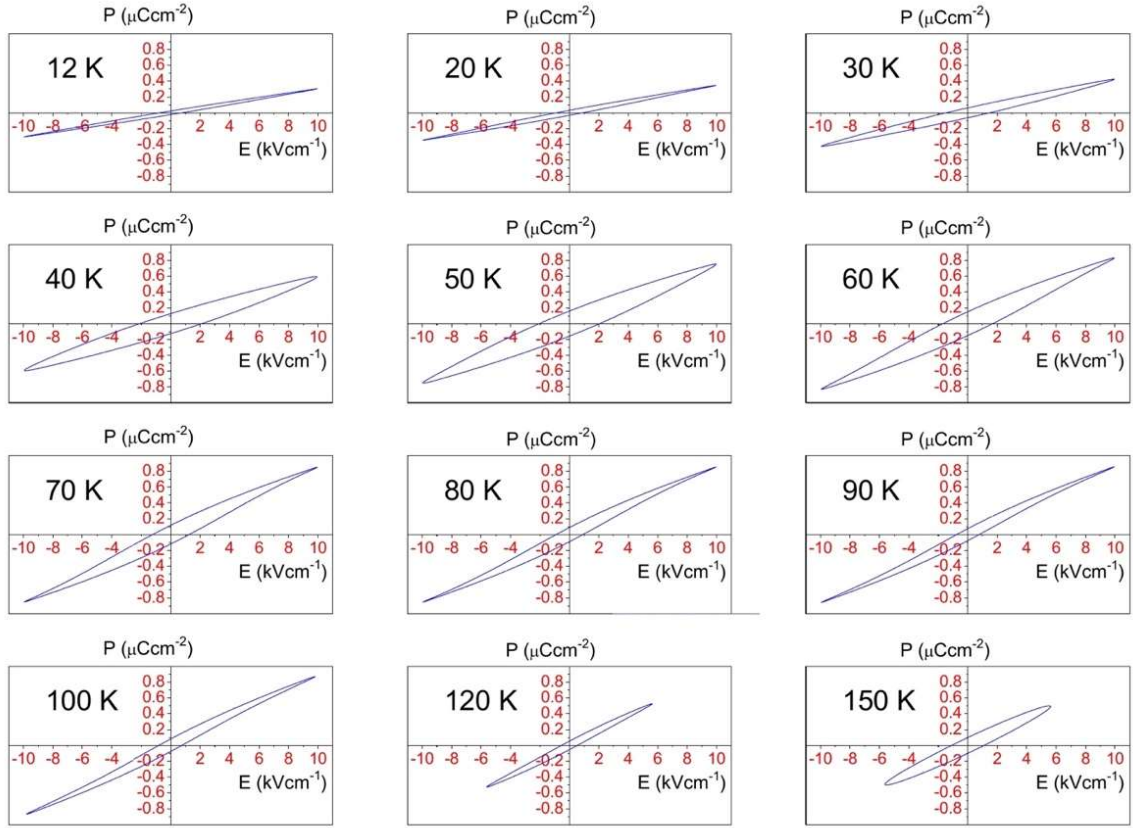

**Figure S8.** *P*(*E*) hysteresis loops at various temperatures taken at 50 Hz. A lossy loop is seen in the paraelectric phase at 150 K, ferroelectric hysteresis loops develop below 120 K. The coercive field, $E_c$, and remnant polarization, $P_r$, increase on cooling from 120 K and then decrease below 50 K, as the maximum applied field of 10 kVcm$^{-1}$ becomes insufficient to induce a single domain state. It was not possible to apply a higher field because the sample becomes leaky. For the same reason, the maximum applied field is lower above 100 K.

In the 30 − 70 K range, both $\epsilon'(T)$ and $\epsilon''(T)$ exhibit a frequency dependent behavior. Such behavior can be explained by a dielectric relaxation with the mean relaxation frequency, $f_R$, corresponding to the frequency-dependent peak in $\epsilon''(T)$ (see inset of Figure S9). The temperature dependence of the relaxation time, $\tau = 1/2\pi f_R$, in BiMn$_3$Cr$_4$O$_{12}$ is shown in Figure S9. The relaxation time follows the Arrhenius law, $\tau(T) = \tau_L exp(U/k_B T)$, where $\tau_L$ is a pre-exponential factor and $U$ is the activation energy. The relaxation describes most probably the thermally activated motion of ferroelectric domain walls in the ferroelectric phase.[7] The activation energy, $U \approx 73$ meV, has a similar value to what has been reported for other ferroelectrics.[7-9]

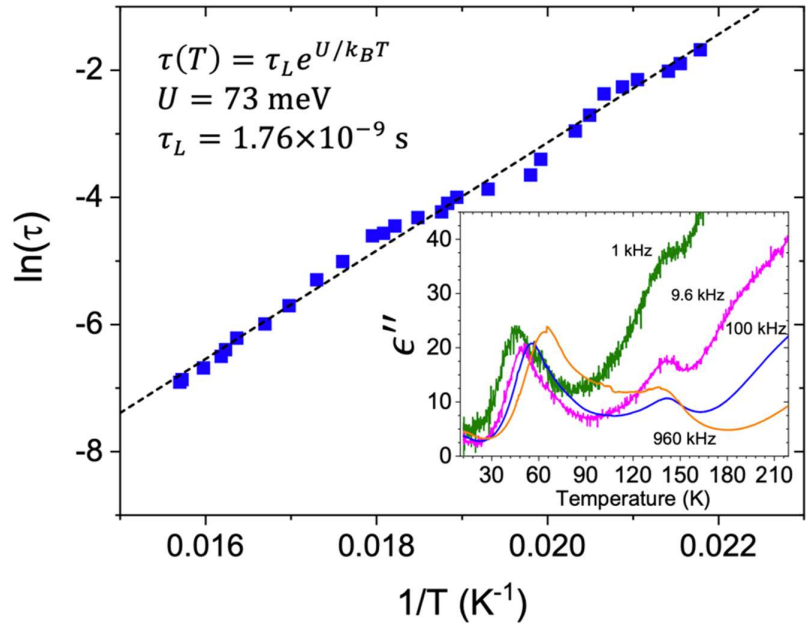

**Figure S9.** Temperature dependence of the relaxation time, $\tau$, in BiMn$_3$Cr$_4$O$_{12}$. The fit with the Arrhenius law is shown as a dashed line. Inset: temperature dependence of the imaginary part of the permittivity measured at selected frequencies.

**Infrared Spectra**

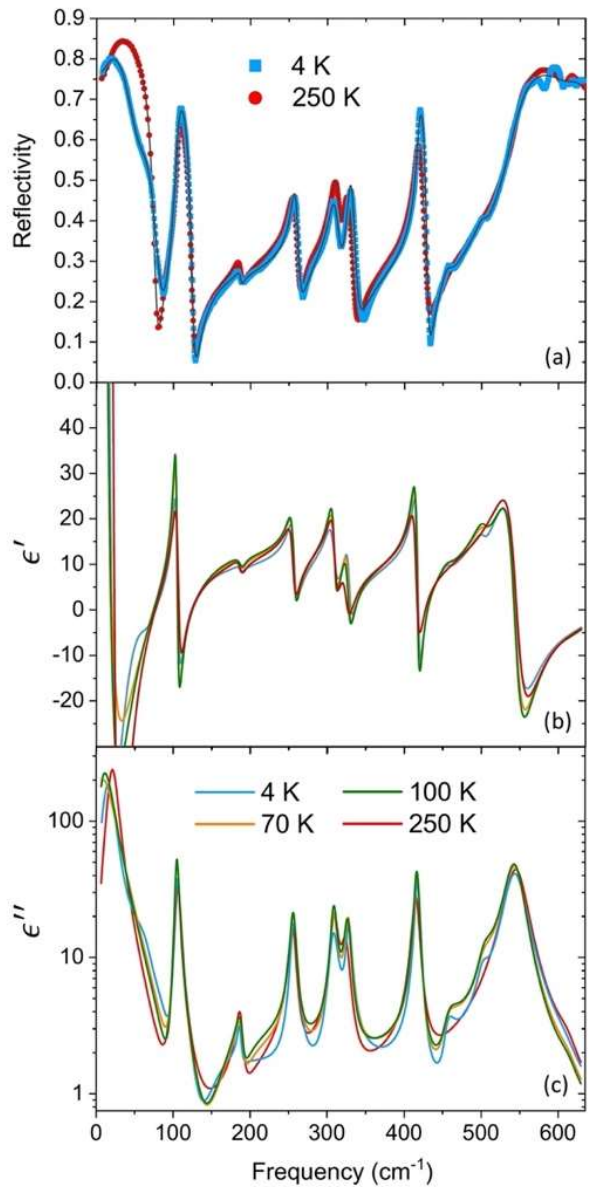

**Figure S10.** (a) IR reflectivity spectra of BiMn$_3$Cr$_4$O$_{12}$ at 4 K and 250 K. The fits of Equation 4 to the reflectivity spectra are shown as dotted lines. (b) $\epsilon'$ and (c) $\epsilon''$ (in logarithmic scale) spectra obtained from fitting the IR reflectivity at several temperatures.

**Table SI.** Parameters of the IR-active modes in BiMn$_3$Cr$_4$O$_{12}$ obtained from the fits of IR reflectivity using Equation 3 and 4. All modes have $F_u$ symmetry in the paraelectric phase. The values predicted from first-principles calculations in the paraelectric phase by Dai et al. are also provided.[4] The oscillator strength, $S_j = \Delta\epsilon_j \omega_{TOj}^2$, is in cm$^{-2}$, while all the other parameters are in cm$^{-1}$.

| | 4 K | | | | | 250 K | | | | Theory |
|---|---|---|---|---|---|---|---|---|---|---|
| $\omega_{TO}$ | $\gamma_{TO}$ | $\omega_{LO}$ | $\gamma_{LO}$ | $S\ (\times 10^4)$ | $\omega_{TO}$ | $\gamma_{TO}$ | $\omega_{LO}$ | $\gamma_{LO}$ | $S\ (\times 10^4)$ | $\omega_{TO}$ |
| 18.7 | 21.6 | 85.2 | 21.9 | 6.73 | 22.3 | 16.3 | 77.6 | 9.6 | 8.16 | 30.74 |
| 71.5 | 33.4 | 78 | 26.1 | 3.03 | 72 | 182.9 | 71.6 | 200 | 0.02 | – |
| 78.5 | 70.3 | 51 | 38.4 | 0.16 | – | – | – | – | – | – |
| 105.1 | 7.2 | 125.9 | 7.2 | 4.88 | 106.1 | 8.4 | 125.0 | 9.8 | 2.48 | 106.28 |
| 186.7 | 5.9 | 187.1 | 5.6 | 0.05 | 187.5 | 10.0 | 188.8 | 8.8 | 0.28 | 180.70 |
| 256.7 | 9.0 | 263.3 | 10.1 | 3.06 | 255.5 | 10.4 | 261.6 | 9.5 | 3.10 | 250.24 |
| 308.1 | 13.4 | 315.1 | 15.1 | 4.95 | 309.6 | 10.6 | 314.9 | 11.9 | 4.65 | 300.01 |
| 328.2 | 8.6 | 337.5 | 13.7 | 3.55 | 324.5 | 13.1 | 334.5 | 11.6 | 3.90 | 313.76 |
| 417.7 | 7.4 | 430.6 | 6.7 | 10.51 | 415.5 | 10.6 | 428.5 | 11.5 | 10.34 | 416.47 |
| 456.0 | 15.2 | 457.0 | 17.6 | 0.76 | – | – | – | – | – | 425.12 |
| 503.1 | 17.2 | 504.5 | 17.4 | 2.70 | – | – | – | – | – | – |
| 544.0 | 32.6 | 614.0 | 39.9 | 68.70 | 545.7 | 33.0 | 611.0 | 35.1 | 71.85 | 574.50 |
| 614.4 | 38.0 | 676.5 | 156.0 | 0.24 | 612.5 | 33.6 | 676.5 | 156 | 1.04 | 590.81 |
| 686.0 | 144.0 | 701.5 | 19.5 | 0.71 | 686.0 | 144.0 | 701.5 | 19.0 | 0.74 | 614.86 |

# Raman Spectra

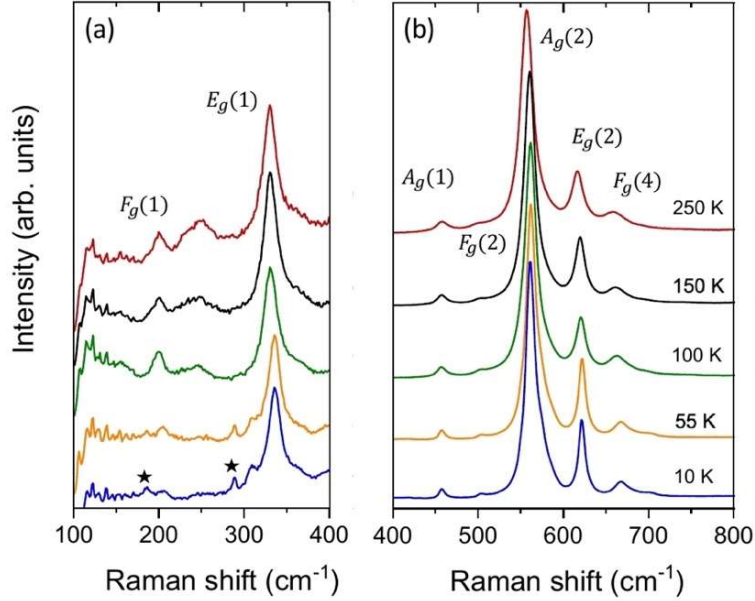

**Figure S11.** Unpolarized Raman spectra of BiMn$_3$Cr$_4$O$_{12}$ measured at several temperatures in the (a) $100-400$ cm$^{-1}$ and (b) $400-800$ cm$^{-1}$ spectral range. The spectra have been shifted and the vertical scales in panels (a) and (b) are different for clarity. The $F_g(3)$ mode near 570 cm$^{-1}$, is screened by the $A_g(2)$ mode. The modes marked by a star arise from BiCrO$_3$ impurities.[10]

**Table SII.** Frequencies of the Raman-active modes in BiMn$_3$Cr$_4$O$_{12}$. The symmetries in the paraelectric phase are indicated. The values predicted from first-principles calculations in the paraelectric phase by Dai et al. are also provided.[4] All values are expressed in cm$^{-1}$.

| Symmetry ($T > T_{FE1}$) | 10 K | 250 K | Theory |
| --- | --- | --- | --- |
| $F_g(1)$ | 199.8 | 199.7 | 181.59 |
| ? | 246 | 243.5 | – |
| $E_g(1)$ | 331.6 | 330.2 | 309.88 |
| $A_g(1)$ | 459.7 | 458.2 | 428.45 |
| $F_g(2)$ | 494.2 | 495.8 | 513.94 |
| $A_g(2)$ | 564.4 | 564.4 | 556.35 |
| $F_g(3)$ | 570.0 | – | 576.26 |
| $E_g(2)$ | 624.5 | 622.9 | 613.45 |
| $F_g(4)$ | 664.1 | 661.2 | 663.02 |